\documentclass[man]{apa7}

\usepackage[american]{babel}
\usepackage{hyperref}
\usepackage{subfigure}
\usepackage{csquotes}
\usepackage[style=apa,sortcites=true,sorting=nyt,backend=biber]{biblatex}
\DeclareLanguageMapping{american}{american-apa}
\title{Computation by Convective Logic Gates and Thermal Communication}
\shorttitle{Convective Computation}

\author{Stuart Bartlett $^{1,2,}$*, Andrew K Gao$^{1,3}$, Yuk L Yung$^1$}

\affiliation{
$^{1}$ Division of Geological and Planetary Sciences, California Institute of Technology, Pasadena, CA 91125, United States\\
$^{2}$ Earth-Life Science Institute, Tokyo Institute of Technology, Tokyo 152-8550, Japan\\
$^{3}$ Yuanpei College, Peking University, Beijing, China}

\leftheader{Bartlett et al.}
\journal{Artificial Life}

\abstract{We demonstrate a novel computational architecture based on fluid convection logic gates and heat flux-mediated information flows. Our previous work demonstrated that Boolean logic operations can be performed by thermally-driven convection flows. In this work, we {use numerical simulations to} demonstrate a different, but universal Boolean logic operation (NOR), performed by simpler convective gates. The gates in the present work do not rely on obstacle flows or periodic boundary conditions, a significant improvement in terms of experimental realizability. Conductive heat transfer links can be used to connect the convective gates, and we demonstrate this with the example of binary half addition. These {simulated} circuits could be constructed in an experimental setting {with modern, 2-dimensional fluidics equipment, such as a thin layer of fluid between acrylic plates. The presented approach thus} introduces a new realm of unconventional, thermal fluid-based computation.}


\authornote{
\addORCIDlink{Stuart Bartlett}{0000-0001-5680-476X}\\
{*Corresponding author, email: \href{mailto:sjbart@caltech.edu}{sjbart@caltech.edu}}}

\begin{document}
\maketitle

\section{Introduction}
Spontaneous convective motion of a heated fluid is arguably the most paradigmatic example of pattern formation and self-organization in a non-equilibrium system. It has been enticing scientists and lay people alike for hundreds of years, and continues to harbor mysteries to this day \parencite{couston2018order,holyst2019flux}. Given its exotic nonlinear dynamics, it is natural to consider the ways in which convective flows process information or compute.\par
The vast majority of modern computing is performed by semiconductor devices and electron flows. However, there is a broad array of non-electronic systems that process information, and such systems have a long history (especially the biological ones). Examples include gene regulatory networks \parencite{benenson,tagkopoulos,macia,weiss}, cellular automata \parencite{beer,crutchfield,feldman,langton,lizier}, chemical reaction networks \parencite{adamatzky,banzhaf,blount,hjelmfelt,magnasco,soloveichik}, computation by the interference of physical waves or concentration profiles \parencite{adamatzky,Kim,steinbock,toth}, nonlinear dynamical systems \parencite{ditto,ditto2,kia}, among other exotic methods \parencite[e.g.,][]{bandyopadhyay,torrejon}.\par
Fluid-based computing has existed for many decades \parencite{adamatzky,avery,bauer,chapline,Foster,gehring,gobhai,levesque,norwood,norwood2,phillips,zilberfarb}, but became largely obsolete in the face of the dramatic advances in electronics and integrated circuits. However, unconventional computing, along with microfluidics, are highly active areas of contemporary research, given the increasing demands for programmable, nanoscale devices that can sense their environments and exhibit adaptive behavior in scenarios that are unfavorable or impractical for electronics.\par 
For example, \cite{Chiu} demonstrated that the parallel nature of a three-dimenisonal microfluidic system can be exploited to solve an NP (non-deterministic polynomial-time) hard problem, and \cite{Thorsen} presented a microfluidic analogue of an integrated circuit with memory functionality akin to random-access memory. Advances in microfluidic memory and control continued with the work of \cite{groisman}, with potential applications in the medical and chemical industries. \cite{Tor} made use of differential flow resistances to perform all of the primary 2-bit Boolean logic operations in a microfluidic system. Droplets are also an important paradigm of fluid computing, and a prominent example of logic operations by such systems was presented by \cite{Cheow}. In the same year, a two-phase bubble system was introduced that was capable of many logic functions including several 2-bit logic gates (sufficient for universal computation), a toggle flip-flop, a ripple counter, timing restoration, a ring oscillator, and an electro-bubble modulator \parencite{Epstein,Prakash}. Interest in the information-processing characteristics of two-phase fluid computation \parencite{draper,katsikis,morgan,Tsompanas} continues to grow apace, as demand for microscale, self-organizing, intelligent systems increases. Note that the examples given above used lithographically-fabricated, isothermal channel systems, in contrast to the 2D thermal cavity systems presented in this work.\par 
Despite the widely explored realm of fluid computation \parencite[see][for an overview]{adamatzky}, the possibility of computation using thermally-driven flows was not considered until our recent work \parencite{Bartlett1,Bartlett2}. In those papers we demonstrat that convective obstacle flows can exhibit bistability, hysteresis, memory, and several Boolean logic operations. However, linking such convective logic gates to form whole circuits would have been challenging.\par
In the present work, we exploit the inherent bistability of 2D convection flows in closed systems with non-integer aspect ratios. These do not require obstacles to exhibit bistability, and also use no-slip vertical walls instead of periodic boundary conditions, thus simplifying the design. {We explore these systems using high accuracy Lattice Boltzmann models, which have been widely used for thermal fluid simulations for several decades \parencite{Bartlett3,Dixit,He,Liu,Pareschi,Peng,succi}. The results of these simulations can} pave the way for implementation in real world settings.\par
The logic operations and cascading ability that we highlight propel convective computation towards Turing-universality \parencite{turing}, a feat that to our knowledge, was not previously demonstrated in a thermally-driven fluid system.\par

\section{Dynamics, Equations of Motion and Simulation Technique for Thermal Convection}
Single phase fluid convection is a prime example of pattern formation in a non-equilibrium system \parencite{Ahlers,Bejan,grossmann,grossmann2,Kays,Manneville,Saltzman}. When the thermal driving force (the dimensionless Rayleigh number) increases above a critical value, the diffusive, static state becomes unstable to perturbations. Fluid parcels that are relatively warm rise and are displaced by colder parcels, which sink. After a transient phase, the system settles into an organized configuration of convection cells.\par
In the present work, we performed a series of numerical simulations of fluid convection systems. We made use of the conventional Boussinesq approximation, which assumes that density variations are sufficiently small to be ignored, except in relation to gravitational body forces. Assuming fluid incompressibility allows the derivation of a simple continuity equation:
\begin{equation}
\nabla\cdot\textbf{v}=0
\end{equation}
where $\textbf{v}=u\hat{i}+w\hat{k}$ is the non-dimensional fluid velocity {($\hat{i}$ and $\hat{k}$ are unit vectors in the horizontal and vertical directions, respectively)}. Conservation of momentum allows the derivation of the following momentum equation:
\begin{equation}
\frac{\partial\textbf{v}}{\partial t}+\textbf{v}\cdot\nabla\textbf{v}+\nabla P=\frac{\nu}{\chi}\nabla^2\textbf{v}+\frac{\nu}{\chi}\frac{\beta g_0\Delta T\delta^3}{\nu\chi}T\hat{k},
\end{equation}
where $P$ is pressure, $\beta$ is the fluid's coefficient of thermal expansion, $g_0$ the acceleration of gravity, $T$ is temperature, {$\Delta T$ is the temperature difference between the upper and lower boundaries,} $\delta$ is the system's vertical size, $\nu$ is the fluid viscosity, and $\chi$ is the fluid thermal diffusivity. From this equation, we can extract the two dimensionless groups that govern convective fluid dynamical behavior: the thermal driving force or Rayleigh number $Ra=\beta g_0\Delta T'\delta^3/\nu\chi$, and the ratio of viscous to thermal diffusivity $Pr=\nu/\chi$. Finally, the advection-diffusion equation describes the transport of internal energy:
\begin{equation}
\frac{\partial T}{\partial t}+\textbf{v}\cdot\nabla T=\nabla^2T.
\end{equation}
The present work made use of a simple computational fluid dynamics technique known as the Lattice Boltzmann Model, which can accurately simulate the thermohydrodynamics of single-phase convection. This method numerically solves the non-equilibrium Boltzmann equation for an incompressible fluid. The momentum equation and the advection-diffusion equation above can then be derived from the Lattice Boltzmann equation \cite{Bartlett3,Dixit,He,Liu,Pareschi,Peng,succi}. The method solves for the instantaneous motion of the fluid and thus can illuminate unsteady and steady state behavior.\par
In this work, we focus on a 2D fluid enclosed by no-slip plates on all four sides. {This is physically equivalent to ordinary solid walls, in contrast to our previous work that used no-slip boundaries on the upper and lower edges, but periodic boundary conditions in the horizontal direction (fluid leaving the right side of the domain returned on the left, and vice versa).} The aspect ratio is 1.5:1, the Rayleigh number is fixed at $Ra=10^4$, and the Prandtl number at $Pr=1$. This Rayleigh number is above the critical value of $Ra_c\approx1706$ to ensure that convection occurs, but sufficiently low that only laminar flows emerge. Fully turbulent flows would require large temperature differences or exotic fluid properties, and would also exhibit disruptive instabilities. The grid size used was 240x360 for each convective gate, with a lattice relaxation time of $\tau_\nu=0.7$ \parencite[see][for further details on this method]{Bartlett3,Peng}.

\section{The Convective NOR Gate}
Our thermal computational system is based upon the universal NOR gate, which has two binary inputs and one output. In order to emulate the truth table of the NOR gate, we use a system that is differentially heated according to the inputs and is also heated in the central $1/9^{th}$ of the lower boundary to a dimensionless temperature of $T_H=1$. The top boundary is kept at a constant dimensionless temperature of $T_C=0$. The two vertical walls and the lower horizontal wall away from the central $1/9^{th}$ are thermally insulated.\par
\begin{figure}[h]
\begin{center}
\subfigure[\label{fig:gate_00}00]{\includegraphics[width=0.45\textwidth,bb=710 0 4540 2370,clip=true]{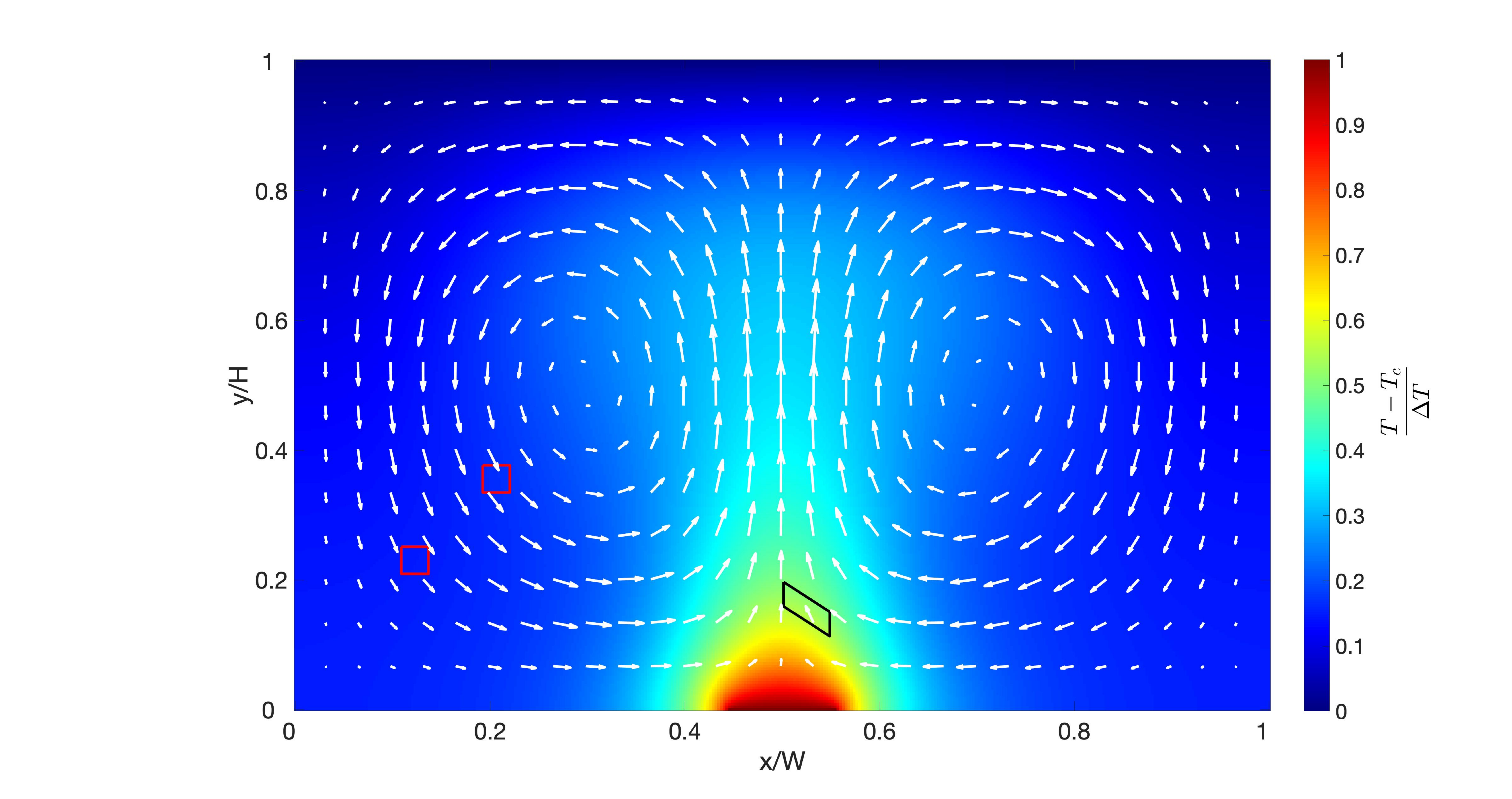}}
\subfigure[\label{fig:gate_01}01]{\includegraphics[width=0.45\textwidth,bb=710 0 4540 2370,clip=true]{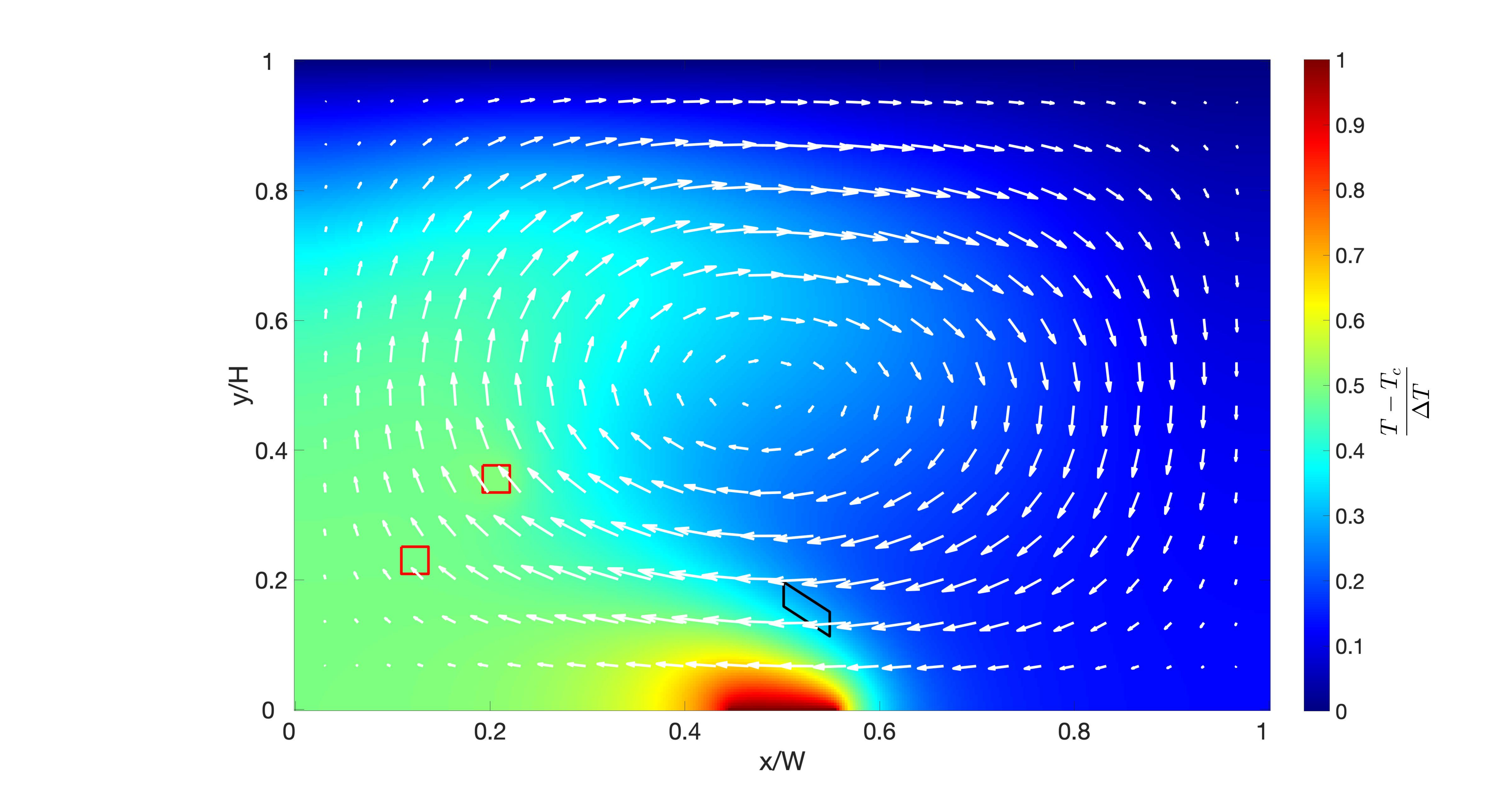}}
\caption{Steady state flow and temperature fields for the convective NOR gate with input a) $(0,0)$, and b) $(0,1)$. Input regions are outlined in red and the output region is outlined in black.}
\label{fig:gate_0}
\end{center}
\end{figure}
\begin{figure}[h]
\begin{center}
\subfigure[\label{fig:gate_10}10]{\includegraphics[width=0.45\textwidth,bb=710 0 4540 2370,clip=true]{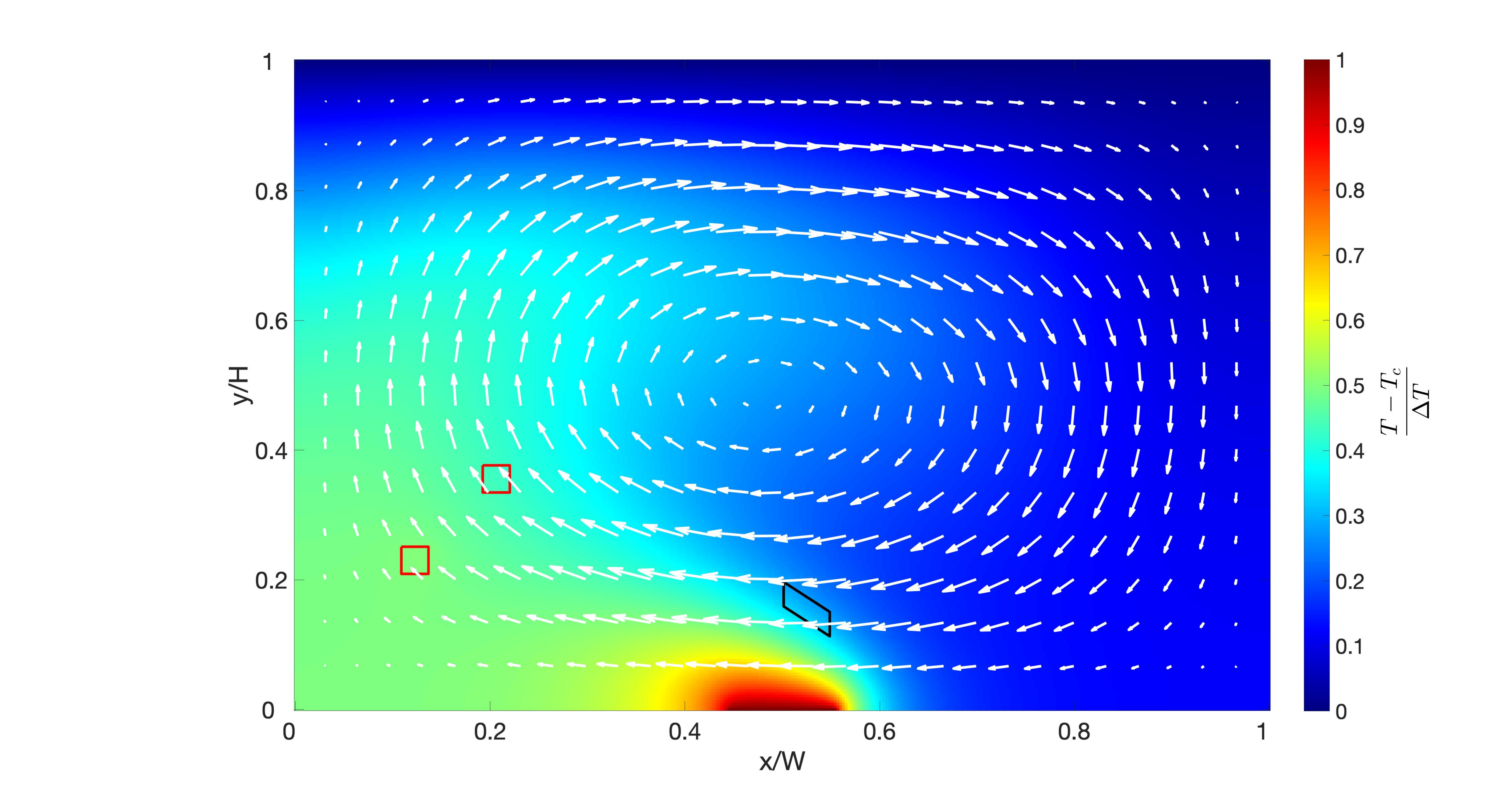}}
\subfigure[\label{fig:gate_11}11]{\includegraphics[width=0.45\textwidth,bb=710 0 4540 2370,clip=true]{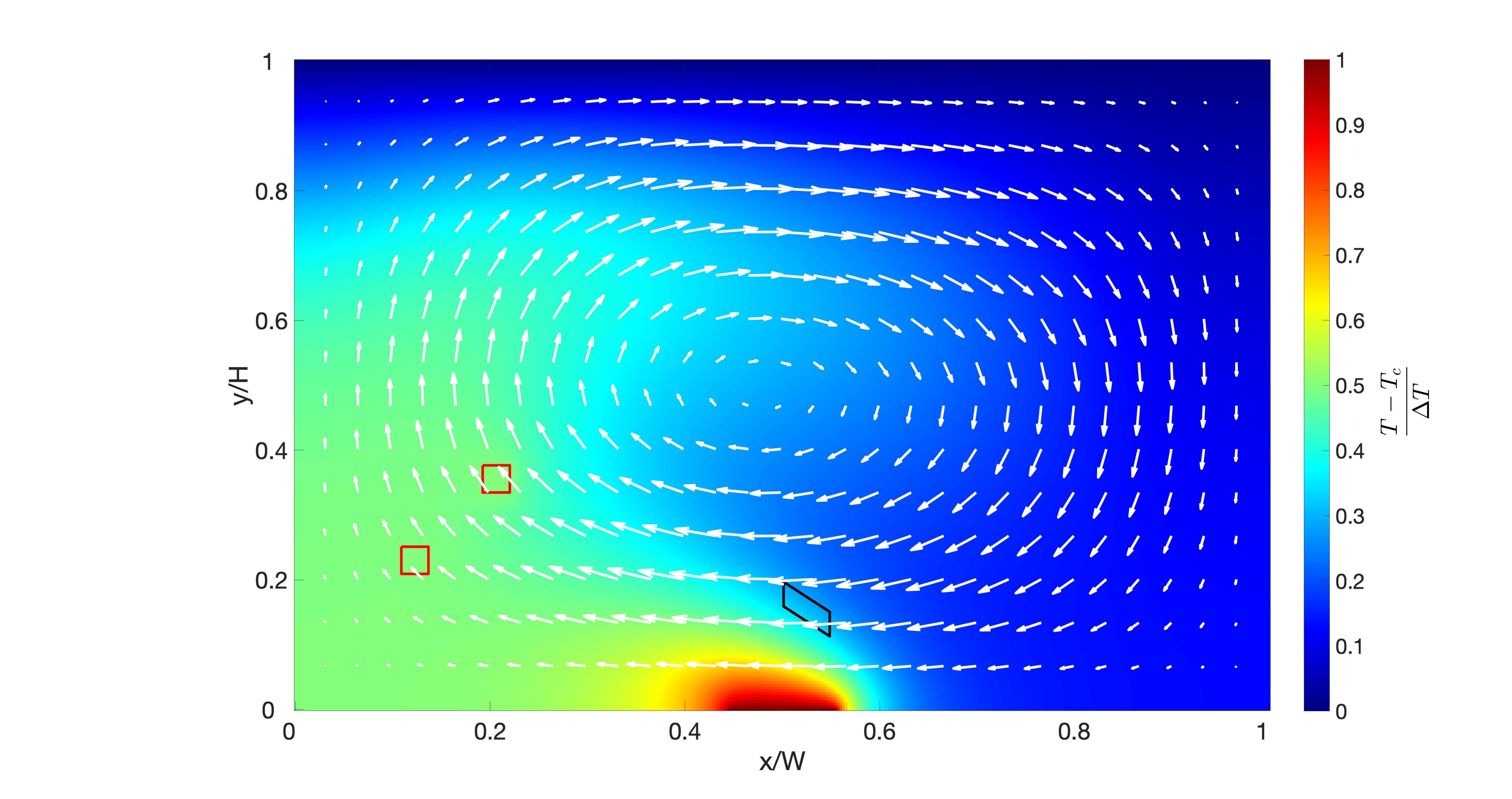}}
\caption{Steady state flow and temperature fields for the convective NOR gate with input a) $(1,0)$, and b) $(1,1)$. Input regions are outlined in red and the output region is outlined in black.}
\label{fig:gate_1}
\end{center}
\end{figure}

The steady states of all four inputs are shown in \autoref{fig:gate_0} and \autoref{fig:gate_1}. \autoref{fig:gate_00} shows the $(0,0)$ input, in which only the central $1/9^{th}$ of the lower boundary is heated. This is due to both inputs being 0, which means there is no heat flux to either of the input regions (an input of 0 corresponds to a complete lack of heat flux to the relevant input region, which also means there is no constraint on the temperature of that input region). The output is represented by the thermal energy of the enclosed region shown in black (inputs shown by the red regions), and is relatively high with this flow field. \autoref{fig:gate_01} shows the $(0,1)$ input, in which the right input region is kept at a dimensionless temperature of 0.5 (we use this temperature instead of 1 because when gates are cascaded the temperature downstream of an `on' gate is $\approx0.5$). Because one of the two inputs is on in this case, the flow field switches from a double convection cell (as in \autoref{fig:gate_00}) to a single convection cell. The aspect ratio of 1.5:1 ensures this bistability. With a 1:1 aspect ratio the most stable state is a single convection cell, and with 2:1 a double convection cell arrangement. With the intermediate value of 1.5:1, at the laminar Rayleigh number used in the present work ($Ra=10^4$), the system is bistable; it can settle into the single or double convection cell state depending on initial conditions and perturbations.\par
\autoref{fig:gate_10} shows the $(1,0)$ input, in which the left input region is heated to dimensionless temperature 0.5. As in the previous case, the stable configuration is the single-cell state. The final input, $(1,1)$, also exhibits this stable state, as shown in \autoref{fig:gate_11}. Having exploited the bistability of this convective system, we see that it can reproduce the input-output relationships of a NOR gate, since the output is low or off for all inputs except $(0,0)$, for which the output is high (on).\par
{\autoref{tab:NOR_tt} summarizes the input-output relationship of our convective NOR gate. The output is given in terms of the average temperature of the output region and the corresponding binary value. This assignment is based on a threshold output temperature of $T_d=0.375$ (see \autoref{sec:circs}).} The next section will consider circuits composed of these gates.\par
\begin{table}[h]
    \centering
    \begin{tabular}{| l | l | l | l |}
    \hline
    Input A & Input B & Output (T) & Output (binary) \\ \hline
    0 & 0 & 0.530 & 1\\ \hline
    0 & 1 & 0.308 & 0\\ \hline
    1 & 0 & 0.303 & 0\\ \hline
    1 & 1 & 0.309 & 0\\ \hline
    \end{tabular}
    \caption{Input-output table for convective NOR gate.}
    \label{tab:NOR_tt}
\end{table}

\section{Convective Logic Circuits}
\label{sec:circs}
In order to cascade our NOR gates into circuits, there must be a means of information transfer between them. We used simple conductive links to achieve this. When connecting gates, we simply assume there is conductive heat flow between the output of the upstream gate and the relevant input of the downstream gate. This could be achieved using metal wires with high thermal conductivity, for example. We parameterized the heat flow through these links with a single parameter representing the thermal conductivity of the link {(the rate at which heat can flow between the input-output regions at either end of the link)}. We do not consider the finite heat capacity {or physical size} of the links, since we assume they are of low mass and volume, and we are only concerned with steady state properties, rather than transient behavior. An additional layer of modelling could be added to simulate the spatial and time-dependent heat flow through the thermal links. However this would likely only reveal the time required for transient effects to vanish, rather than any additional physical effects. Such an extra modelling step could be introduced in preparation for experimental realization, so as to help estimate the required physical parameters.\par
The heat conductance parameter was fixed at 0.05 (a value of 0 would imply complete thermal resistance and a value of 1 would imply no resistance or perfect thermal conduction). This value was chosen through experimentation and found to strike a reasonable balance between ease of information transfer, and suppression of unwanted and transient heat flows. Thermal resistances that were excessively high prevented information flow through the thermal links. In contrast, very low thermal resistances allowed pathological, upstream information transfer. The chosen value of 0.05 reflects the optimal trade-off found between these effects.\par
Note that the gates are assumed to be spatially separated and hence completely thermally isolated from one another (apart from via the conductive links). Our initial designs used heat exchange via shared horizontal boundaries between the gates, but this method proved insufficient to appropriately transfer information between gates.\par
\begin{figure}[h]
\begin{center}
\subfigure[\label{fig:circ_00}00]{\includegraphics[width=0.8\textwidth,bb=240 100 1750 1070,clip=true]{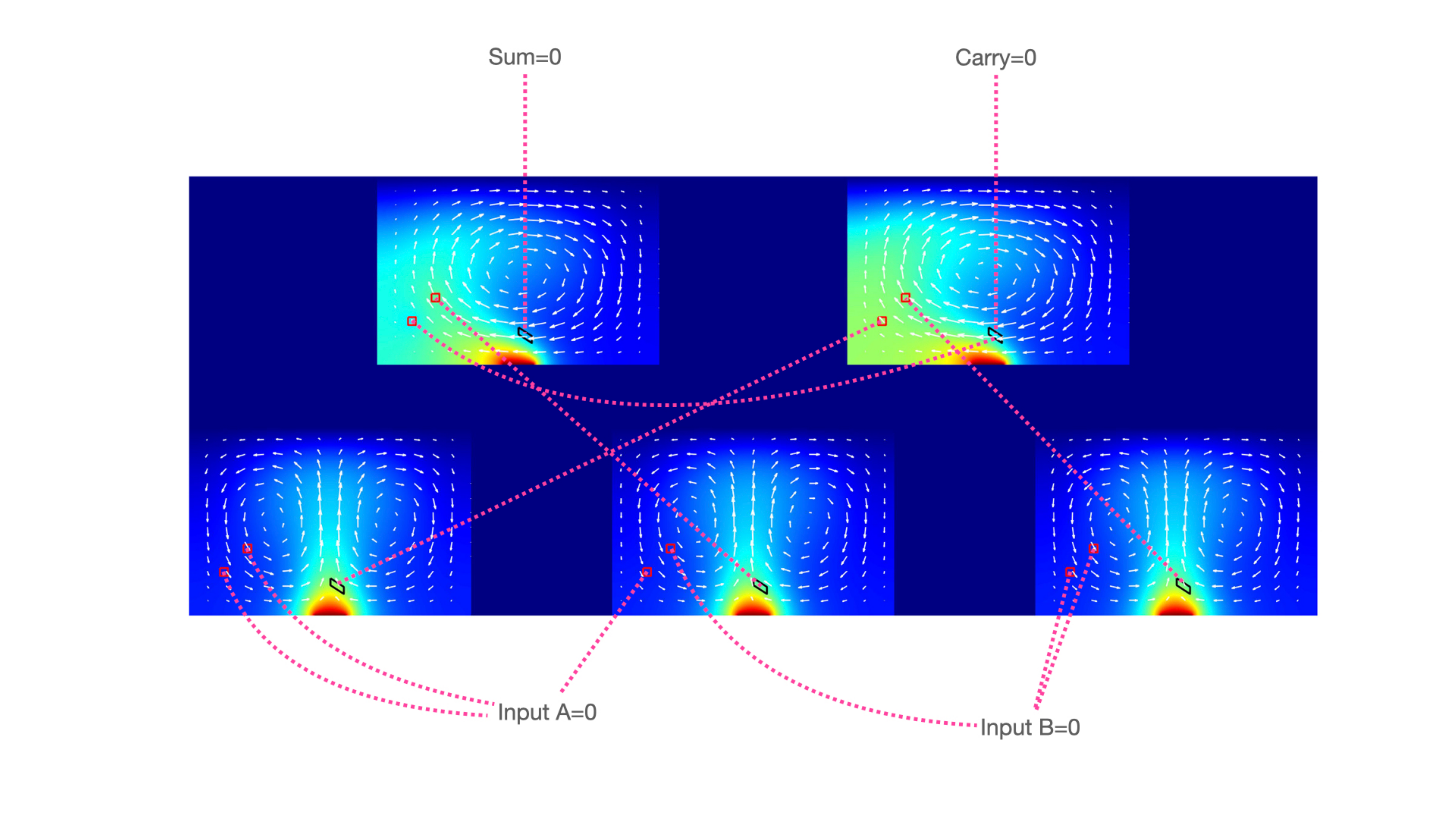}}
\subfigure[\label{fig:circ_01}01]{\includegraphics[width=0.8\textwidth,bb=240 100 1750 1070,clip=true]{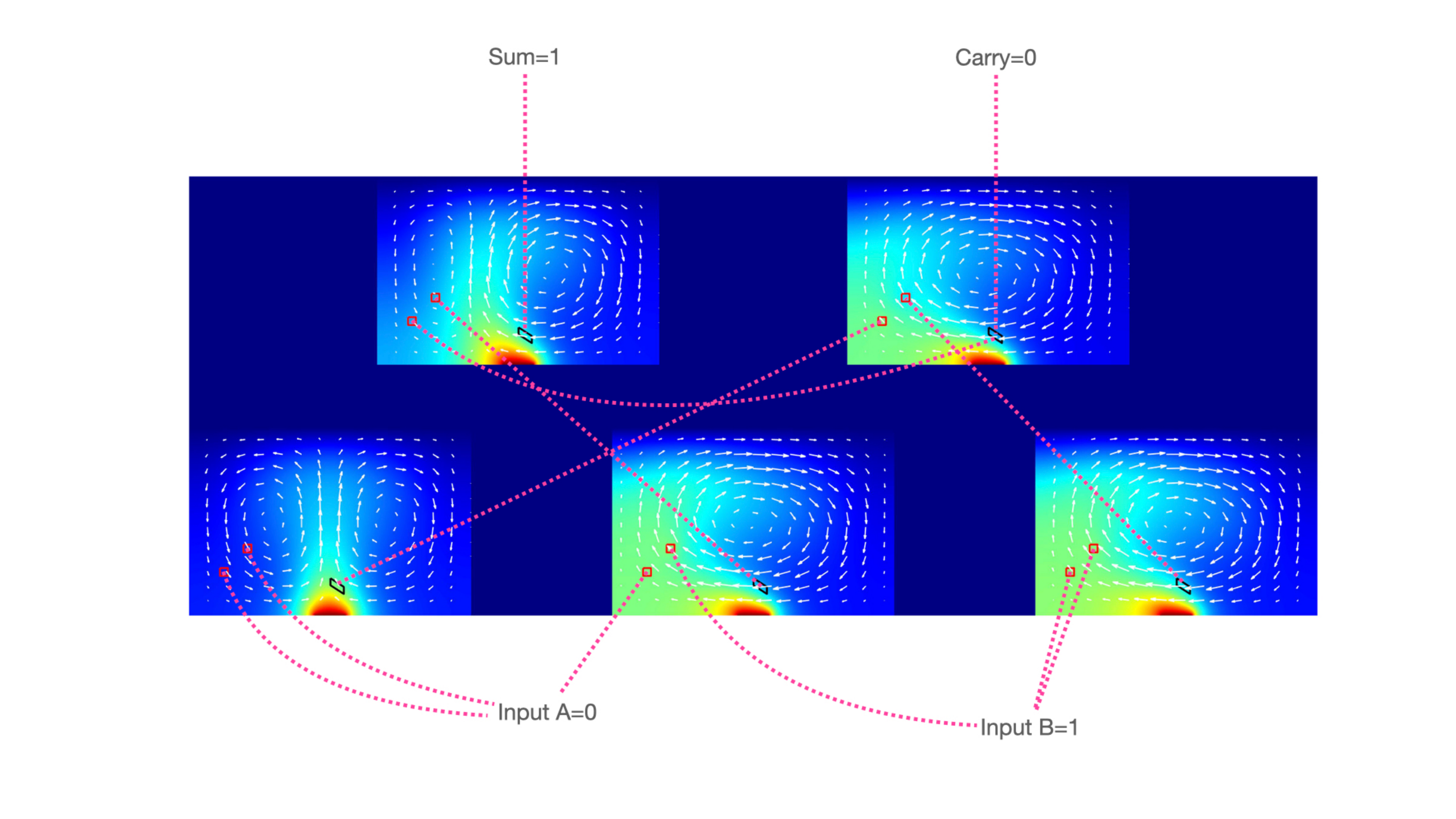}}
\caption{Steady state flow and temperature fields for a binary half addition circuit comprised of five convective NOR gates, with input a) $(0,0)$, and b) $(0,1)$. Input regions are outlined in red and output regions are outlined in black.}
\label{fig:circ_0}
\end{center}
\end{figure}
\autoref{fig:circ_00} shows the steady state configuration of our convective half addition circuit with input $(0,0)$. Conductive links as well as input and output connections are shown with dotted mauve lines. The lower left and lower right gates provide high heat fluxes to the upper right gate via their respective links, causing it to settle into the single convection cell state. The lower three gates remain in the two-cell state. The heat flux from the lower middle gate to the upper left gate causes it to stabilize in the single-cell state. There is little heat flux from the upper right to upper left gate. The resulting outputs are low temperatures in the output regions of the upper gates, corresponding to sum and carry values of 0 and 0.\par
\autoref{fig:circ_01} shows the steady state configuration of our convective half addition circuit with input $(0,1)$. In this case, heat from the lower left gate causes the upper right gate to settle into the single cell state. Because the lower middle and upper right gates are in the single-cell state, there is insufficient heat flux to perturb the upper left gate away from the double convection cell state. Hence the output in this case is $sum=1$ and $carry=0$, as required.\par
\begin{figure}[h]
\begin{center}
\subfigure[\label{fig:circ_10}10]{\includegraphics[width=0.8\textwidth,bb=240 100 1750 1070,clip=true]{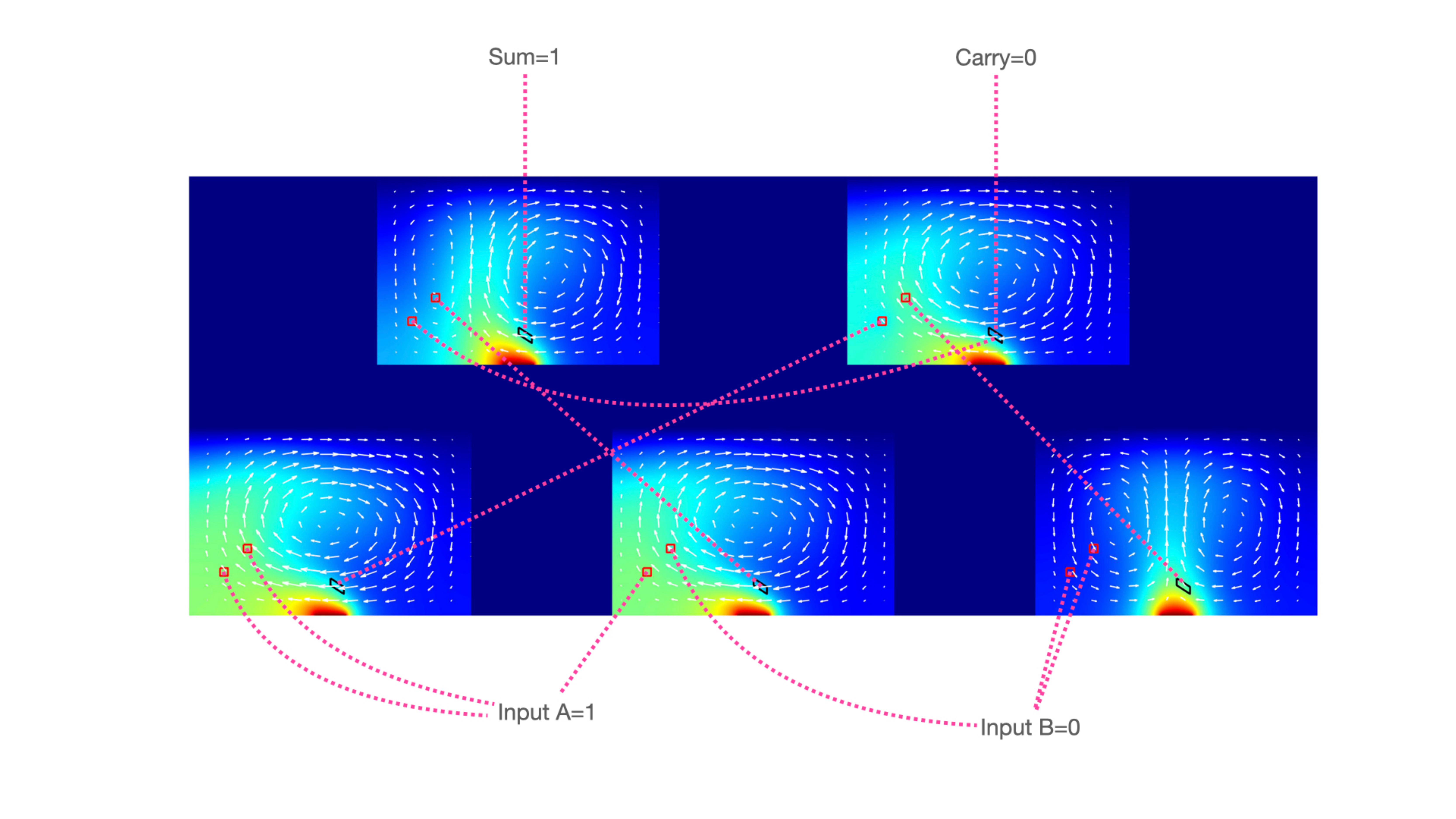}}
\subfigure[\label{fig:circ_11}11]{\includegraphics[width=0.8\textwidth,bb=240 100 1750 1070,clip=true]{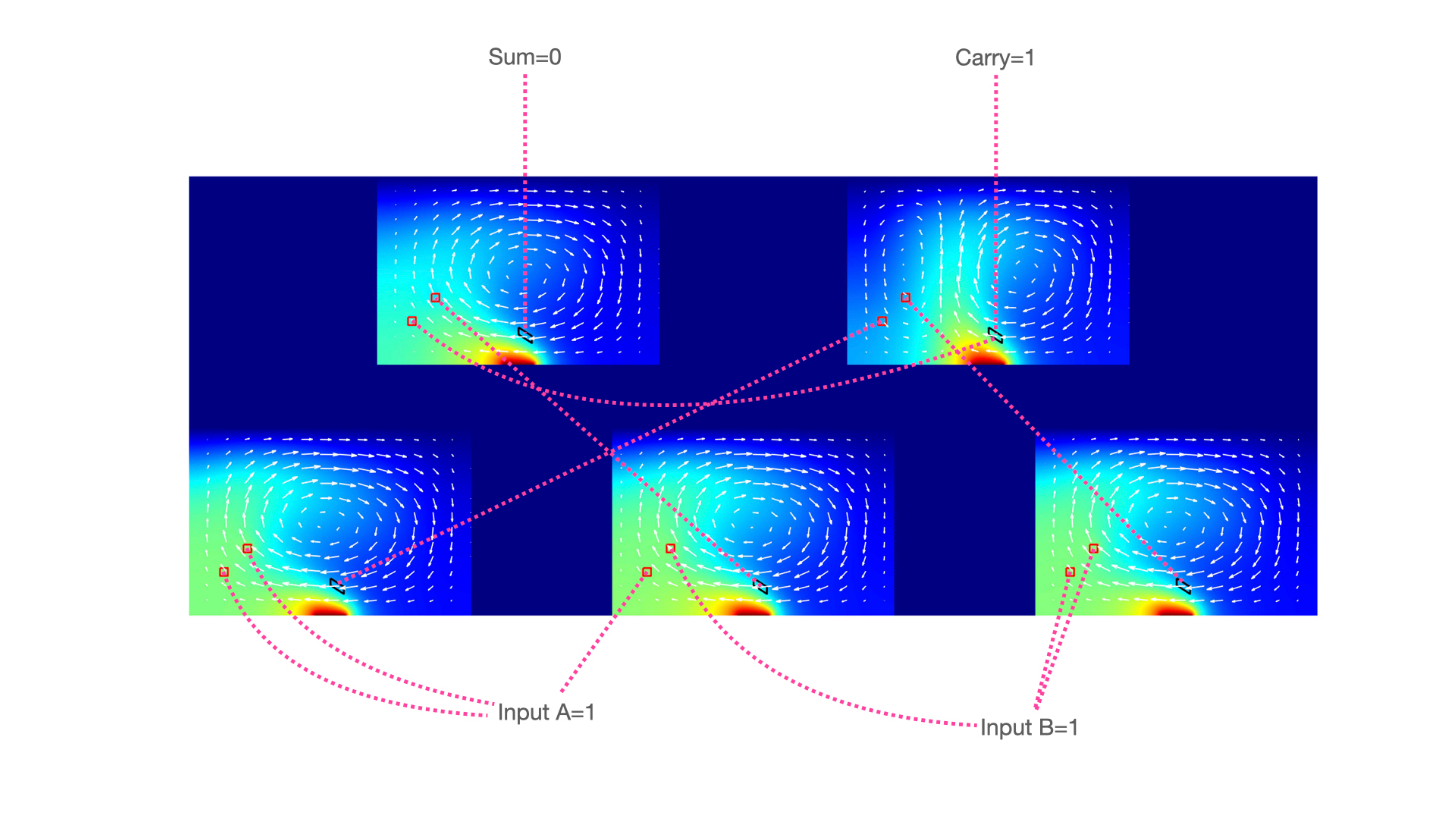}}
\caption{Steady state flow and temperature fields for a binary half addition circuit comprised of five convective NOR gates, with input a) $(1,0)$, and b) $(1,1)$. Input regions are outlined in red and output regions are outlined in black.}
\label{fig:circ_1}
\end{center}
\end{figure}
\autoref{fig:circ_10} shows the steady state configuration of our half addition circuit with input $(1,0)$. The lower left and lower middle gates are in the single-cell state due to the input heat fluxes. The lower right gate remains in the double-cell state and heating from its output provokes the upper right gate to settle into the single-cell state. Due to low heating from the lower middle and upper right gate outputs, the upper left gate remains in the double-cell state, although it is slightly perturbed due to the small heat flux from the two upstream gates. The output in this case is again $sum=1$, $carry=0$.\par
\autoref{fig:circ_11} shows the steady state configuration of our convective NOR circuit with input $(1,1)$. All three lower gates are now in the single-cell state due to the input heat fluxes. This means there is insufficient heat flux to the upper right gate to perturb it into the single-cell state. Since the upper right gate remains in the double-cell state, its output provides heat to the first input of the upper left gate, which settles into the single-cell state and produces a low output. Thus, the output in this case is $sum=1$, $carry=1$. Hence all four inputs produce the correct output according to the binary half addition truth table as shown in \autoref{tab:bin_add_tt}. The third and fourth columns of the table show the average temperatures of the output regions for the sum and carry gates. The feature which distinguishes the states of our gates is whether they have one or two convection cells. We can also assign a threshold output temperature that distinguishes the binary output state. A value of $T_d=0.375$ appropriately delineates between values corresponding to the two binary output states.\par
\begin{table}[h]
    \centering
    \begin{tabular}{| l | l | l | l | l | l |}
    \hline
    Input A & Input B & Sum (T) & Carry (T) & Sum (binary) & Carry (binary) \\ \hline
    0 & 0 & 0.310 & 0.357 & 0 & 0\\ \hline
    0 & 1 & 0.388 & 0.292 & 1 & 0\\ \hline
    1 & 0 & 0.388 & 0.299 & 1 & 0\\ \hline
    1 & 1 & 0.307 & 0.412 & 0 & 1\\ \hline
    \end{tabular}
    \caption{Input-output table for convective binary half addition circuit.}
    \label{tab:bin_add_tt}
\end{table}
The final configuration of our NOR gate resulted from a series of incremental design modifications. The slanted shape of the output, placed slightly to the right of the center of the gate, is configured to maximally reflect the temperature differences due to the two different flow configurations. With a symmetric, central output region, there was undesired heating of the output under the single cell state (which must produce a low thermal output). The slanted shape of the output is designed to further avoid spurious output heating under the single convection cell state.\par
The positioning of the input regions was designed to maximize the influence of input heating on the flow field. They were placed in a transverse arrangement relative to the flow field in order to minimize thermal cross-talk due to advective heat transport (when placed in line with the flow field, excessive heat transfer could occur between them).\par
The positioning of the input and output regions was crucial to enable correct flipping between the two flow configurations, while avoiding spurious information transfer. Given the symmetric nature of conductive heat flux, avoiding upstream thermal influences was particularly difficult. While the present design achieves the correct function for binary half addition, the next phase of this work will develop the design further such that it is more robust, and scalable to larger circuits. This may require additional components to achieve signal restoration and amplification. Furthermore, a material that could act as a thermal diode within the links (allowing heat flow in a given direction but not the reverse), would solve the aforementioned issues related to unwanted upstream information flow.\par
We envisage that our system could be constructed using modern fluidics apparatus, by holding small volumes of fluid between acrylic plates such that their motion is effectively 2-dimensional. The size of the gates and fluid properties would be adjusted to achieve the desired Rayleigh number. The input and output regions would be constructed using sections of thermally conductive material (e.g. metal such as copper, or perhaps more exotic materials with high thermal conductivity), embedded in the acrylic walls of the gates so as not to disturb the hydrodynamics (except through their transport of heat). The links between gates would also be required to be of high thermal conductivity and minimal heat capacity. However, the conductivity would likely require tuning to achieve the desired input-output behavior, and the fact that any real material has a finite heat capacity may also influence the performance of the gates. At the very least it would introduce a delay in signal propagation compared to our numerical simulations. Hence there would be a degree of engineering required to achieve the performance demonstrated in this work, but given the vast accomplishments of microfluidics and fluidic computation in recent years, it seems highly likely that all the relevant issues can be resolved.\par

\section{Conclusions}
We have presented a novel natural computing system based on convective fluid logic operations, and thermal information transfer. The `convective NOR gate' was shown to reproduce the input-output relationship of the standard NOR gate, using the bistability of laminar fluid convection. This bistability arises in systems with non-integer aspect ratio, since two states (a single convection cell or two convection cells) are similarly stable. Such convective gates can be cascaded into circuits using conductive heat transfer connections. We demonstrated the function of a complete circuit using five gates and four links, that emulated a binary half addition circuit. All four inputs produced the correct outputs, supporting the possibility that thermal fluid computation has the potential to be developed into a Turing-universal logic system.\par
{Modern fluidics apparatus should be capable of implementing the required thermohydrodynamics of our system, for example using thin sections of fluid held between acrylic plates. Input-output regions and conductive links would have to be carefully engineered to exhibit the required thermal conductivities and low heat capacities. Tuning of the sizes and relevant physical parameters would be required, but this engineering process would likely garner new insights into these networked convective gates and may even inspire novel and more effective designs.}

\section*{Acknowledgments}
This work was supported in part by the Caltech Division of Geological and Planetary Sciences Discovery Fund. Y. L. Y. was supported in part by the NAI Virtual Planetary Laboratory grant at the University of Washington.

\section*{Data Availability}
The data that support the findings of this study are available from the corresponding author upon reasonable request.

\printbibliography
\end{document}